\documentclass[12pt]{article}

\usepackage[margin=1in,dvips,a4paper,nohead]{geometry}
\usepackage[margin=1em, font=small, justification=centerlast, tableposition=top]{caption}
\usepackage{float}

\usepackage{amsfonts,amssymb,amsmath}
\allowdisplaybreaks

\usepackage[noBBpl,slantedGreek]{mathpazo}

\usepackage{graphicx}

\usepackage{pstricks,pst-node}
\psset{arrowscale=1.5}

\def\orientedEdge(#1,#2)(#3,#4){%
  \pcline(#1,#2)(#3,#4)%
  \lput{:U}{
    \pspicture(0,0)(0,0)
     \psline[arrows=->](0,0)(1.8pt,0)
    \endpspicture
  }
}

\def \bbZ {\mathbb Z}
\def \rme {\mathrm e}
\def \rmi {\mathrm i}

\begin{document}

\title{Three-coloring statistical model with\\ domain wall boundary conditions.\\ I. Functional equations}
\author{A.~V.~Razumov, Yu.~G.~Stroganov\\
\small \it Institute for High Energy Physics\\[-.5em]
\small \it 142281 Protvino, Moscow region, Russia}
\date{}

\maketitle

\begin{abstract}
In 1970 Baxter considered the statistical three-coloring lattice model for the case of toroidal boundary conditions. He used the Bethe ansatz and found the partition function of the model in the thermodynamic limit. We consider the same model but use other boundary conditions for which one can prove that the partition function satisfies some functional equations similar to the functional equations satisfied by the partition function of the six-vertex model for a special value of the crossing parameter.
\end{abstract}

\section{Introduction}
\label{s:1}

We consider various colorings of a square $n \times m$ lattice with three colors, such that any two adjacent faces have different colors. Sometimes one imposes some boundary conditions restricting possible colorings. The most common case here is the toroidal boundary conditions, when the colors  of the first face must be different from the color of the last one for each row and column of the lattice. An example is given in Figure~\ref{ThreeColoringExample},
\begin{figure}[htb]
\centering
\begin{pspicture}(-.1,-.1)(6.1,5.1)
\multips(0,0)(0,1){6}{
  \psline(0,0)(6,0)}
\multips(0,0)(1,0){7}{
  \psline(0,0)(0,5)  }
\multips(0,0)(0,1){6}{
  \multips(0,0)(1,0){7}{
    \pscircle[fillstyle=solid](0,0){.08}}}
\rput(.5,.5){$\scriptstyle \overline 2$}
\rput(1.5,.5){$\scriptstyle \overline 1$}
\rput(2.5,.5){$\scriptstyle \overline 2$}
\rput(3.5,.5){$\scriptstyle \overline 1$}
\rput(4.5,.5){$\scriptstyle \overline 0$}
\rput(5.5,.5){$\scriptstyle \overline 1$}
\rput(.5,1.5){$\scriptstyle \overline 0$}
\rput(1.5,1.5){$\scriptstyle \overline 2$}
\rput(2.5,1.5){$\scriptstyle \overline 1$}
\rput(3.5,1.5){$\scriptstyle \overline 2$}
\rput(4.5,1.5){$\scriptstyle \overline 1$}
\rput(5.5,1.5){$\scriptstyle \overline 2$}
\rput(.5,2.5){$\scriptstyle \overline 2$}
\rput(1.5,2.5){$\scriptstyle \overline 1$}
\rput(2.5,2.5){$\scriptstyle \overline 2$}
\rput(3.5,2.5){$\scriptstyle \overline 1$}
\rput(4.5,2.5){$\scriptstyle \overline 0$}
\rput(5.5,2.5){$\scriptstyle \overline 1$}
\rput(.5,3.5){$\scriptstyle \overline 0$}
\rput(1.5,3.5){$\scriptstyle \overline 2$}
\rput(2.5,3.5){$\scriptstyle \overline 0$}
\rput(3.5,3.5){$\scriptstyle \overline 2$}
\rput(4.5,3.5){$\scriptstyle \overline 1$}
\rput(5.5,3.5){$\scriptstyle \overline 2$}
\rput(.5,4.5){$\scriptstyle \overline 1$}
\rput(1.5,4.5){$\scriptstyle \overline 0$}
\rput(2.5,4.5){$\scriptstyle \overline 1$}
\rput(3.5,4.5){$\scriptstyle \overline 0$}
\rput(4.5,4.5){$\scriptstyle \overline 2$}
\rput(5.5,4.5){$\scriptstyle \overline 0$}
\end{pspicture}
\caption{}
\label{ThreeColoringExample}
\end{figure}
where we use a convenient labeling of the colors by the elements of the ring $\bbZ_3$. We denote the elements of $\bbZ_3$ by $\overline 0$, $\overline 1$ and $\overline 2$. When an element of $\bbZ_3$ arises in a context where an integer should be, it is treated as the corresponding integer $0$, $1$ or $2$. Vice verse, when an integer $i$ arises in a context where an element of $\bbZ_3$ should be, it is treated as the element of $\bbZ_3$ corresponding to the reminder after division of $i$ by 3. It is clear that the colors of the adjacent faces are different if and only if the corresponding elements of $\bbZ_3$ differ by $+\overline 1$ or $-\overline 1$.

The simplest combinatorial problem arising here is to enumerate all possible colorings. More generally one can try to find the numbers $C_{n,m}(k_{\overline 0}, k_{\overline 1}, k_{\overline 2} )$ of colorings with $k_{\overline 0}$ faces of color $\overline 0$, $k_{\overline 1}$ faces of color $\overline 1$, and $k_{\overline 2}$ faces of color $\overline 2$. Certainly, the numbers $C_{n,m}(k_{\overline 0}, k_{\overline 1}, k_{\overline 2} )$ are different from zero only if $k_{\overline 0} + k_{\overline 1} + k_{\overline 2} = n m$. It is useful to introduce the generating function
\[
Z_{n, m}(z_{\overline 0}, z_{\overline 1}, z_{\overline 2}) = \sum_{\substack{k_{\overline 0}, k_{\overline 1}, k_{\overline 2} \\ k_{\overline 0} + k_{\overline 1} + k_{\overline 2} = n m}} z^{k_{\overline 0}}_{\overline 0} z^{k_{\overline 1}}_{\overline 1} z^{k_{\overline 2}}_{\overline 2} \, C_{n,m}(k_{\overline 0}, k_{\overline 1}, k_{\overline 2}),
\]
where $z_{\overline 0}, z_{\overline 1}, z_{\overline 2}$ are arbitrary numbers.

One can treat colorings as states of a quantum statistical model regarding colors as states of faces. Here the numbers $z_{\overline 0}, z_{\overline 1}, z_{\overline 2}$ are the corresponding Boltzmann weights of faces, the Boltzmann weight of a state is the product of the Boltzmann weights of the faces, and $Z_{n, m}(z_{\overline 0}, z_{\overline 1}, z_{\overline 2})$ is the partition function (state sum) of the model. It is convenient to represent the weight of a state as the product of the weights of the vertices of the lattice. Here if a vertex belongs to four faces of colors $r_1$, $r_2$, $r_3$, and $r_4$, we assign to it the weight $(z_{r_1} z_{r_2} z_{r_3} z_{r_4})^{1/4}$. One can generalize the model assuming that the weight of a vertex is an arbitrary function of the colors of the four adjacent faces. It is also possible to introduce a dependence on a spectral parameter and satisfy the Yang--Baxter equations (star-triangle relation) to obtain an integrable system.

In 1970 Baxter, using an appropriate version of the Bethe ansatz, found the partition function $Z_{n, m}(z_{\overline 0}, z_{\overline 1}, z_{\overline 2})$ for the toroidal boundary conditions in the thermodynamic limit~\cite{Bax70, Bax82}. After that, one of the authors of the present paper constructed a solution of the Yang--Baxter equations which, for the special value of the spectral parameter, leads to the partition function of the three-coloring model~\cite{Str82}. With the help of the inversion trick~\cite{Str79}, the largest eigenvalues for the family of commuting transfer matrices were found  and the results of the paper~\cite{Bax70} were reproduced.

To describe the solution of the Yang--Baxter equations given in the work~\cite{Str82}, note that there are six types of possible vertex color configurations  given by the first raw of the pictures in Figure~\ref{ThreeColoringToSixVertex},
\begin{figure}[htb]
\centering
\begin{tabular}{cccccc}
\begin{pspicture}(-.1,-.1)(2.1,2.1)
\psline (0,0)(2,0)
\psline (0,1)(2,1)
\psline (0,2)(2,2)
\psline (0,0)(0,2)
\psline (1,0)(1,2)
\psline (2,0)(2,2)
\multips(0,0)(0,1){3}{
  \multips(0,0)(1,0){3}{
    \pscircle[fillstyle=solid](0,0){.08}}}
\rput(.5,1.5){$\scriptstyle r - 1$}
\rput(1.5,1.5){$\scriptstyle r$}
\rput(1.5,.5){$\scriptstyle r + 1$}
\rput(.5,.5){$\scriptstyle r$}
\end{pspicture} &
\begin{pspicture}(-.1,-.1)(2.1,2.1)
\psline (0,0)(2,0)
\psline (0,1)(2,1)
\psline (0,2)(2,2)
\psline (0,0)(0,2)
\psline (1,0)(1,2)
\psline (2,0)(2,2)
\multips(0,0)(0,1){3}{
  \multips(0,0)(1,0){3}{
    \pscircle[fillstyle=solid](0,0){.08}}}
\rput(.5,1.5){$\scriptstyle r + 1$}
\rput(1.5,1.5){$\scriptstyle r$}
\rput(1.5,.5){$\scriptstyle r - 1$}
\rput(.5,.5){$\scriptstyle r$}
\end{pspicture} &
\begin{pspicture}(-.1,-.1)(2.1,2.1)
\psline (0,0)(2,0)
\psline (0,1)(2,1)
\psline (0,2)(2,2)
\psline (0,0)(0,2)
\psline (1,0)(1,2)
\psline (2,0)(2,2)
\multips(0,0)(0,1){3}{
  \multips(0,0)(1,0){3}{
    \pscircle[fillstyle=solid](0,0){.08}}}
\rput(.5,1.5){$\scriptstyle r$}
\rput(1.5,1.5){$\scriptstyle r - 1$}
\rput(1.5,.5){$\scriptstyle r$}
\rput(.5,.5){$\scriptstyle r + 1$}
\end{pspicture} &
\begin{pspicture}(-.1,-.1)(2.1,2.1)
\psline (0,0)(2,0)
\psline (0,1)(2,1)
\psline (0,2)(2,2)
\psline (0,0)(0,2)
\psline (1,0)(1,2)
\psline (2,0)(2,2)
\multips(0,0)(0,1){3}{
  \multips(0,0)(1,0){3}{
    \pscircle[fillstyle=solid](0,0){.08}}}
\rput(.5,1.5){$\scriptstyle r$}
\rput(1.5,1.5){$\scriptstyle r + 1$}
\rput(1.5,.5){$\scriptstyle r$}
\rput(.5,.5){$\scriptstyle r - 1$}
\end{pspicture} &
\begin{pspicture}(-.1,-.1)(2.1,2.1)
\psline (0,0)(2,0)
\psline (0,1)(2,1)
\psline (0,2)(2,2)
\psline (0,0)(0,2)
\psline (1,0)(1,2)
\psline (2,0)(2,2)
\multips(0,0)(0,1){3}{
  \multips(0,0)(1,0){3}{
    \pscircle[fillstyle=solid](0,0){.08}}}
\rput(.5,1.5){$\scriptstyle r$}
\rput(1.5,1.5){$\scriptstyle r + 1$}
\rput(1.5,.5){$\scriptstyle r$}
\rput(.5,.5){$\scriptstyle r + 1$}
\end{pspicture} &
\begin{pspicture}(-.1,-.1)(2.1,2.1)
\psline (0,0)(2,0)
\psline (0,1)(2,1)
\psline (0,2)(2,2)
\psline (0,0)(0,2)
\psline (1,0)(1,2)
\psline (2,0)(2,2)
\multips(0,0)(0,1){3}{
  \multips(0,0)(1,0){3}{
    \pscircle[fillstyle=solid](0,0){.08}}}
\rput(.5,1.5){$\scriptstyle r$}
\rput(1.5,1.5){$\scriptstyle r - 1$}
\rput(1.5,.5){$\scriptstyle r$}
\rput(.5,.5){$\scriptstyle r - 1$}
\end{pspicture} \\[1em]
$\alpha_r(\varphi | \lambda, p)$ & $\alpha'_r(\varphi | \lambda, p)$ & $\beta_r(\varphi | \lambda, p)$ & $\beta'_r(\varphi | \lambda, p)$ & $\gamma_r(\varphi | \lambda, p)$ & $\gamma'_r(\varphi | \lambda, p)$ \\[1em]
\begin{pspicture}(0,0)(2,2)
\orientedEdge(1,1)(1,2)
\orientedEdge(1,1)(2,1)
\orientedEdge(1,0)(1,1)
\orientedEdge(0,1)(1,1)
\pscircle[fillstyle=solid](1,1){.6mm}
\end{pspicture} &
\begin{pspicture}(0,0)(2,2)
\orientedEdge(1,2)(1,1)
\orientedEdge(2,1)(1,1)
\orientedEdge(1,1)(1,0)
\orientedEdge(1,1)(0,1)
\pscircle[fillstyle=solid](1,1){.6mm}
\end{pspicture} &
\begin{pspicture}(0,0)(2,2)
\orientedEdge(1,2)(1,1)
\orientedEdge(1,1)(2,1)
\orientedEdge(1,1)(1,0)
\orientedEdge(0,1)(1,1)
\pscircle[fillstyle=solid](1,1){.6mm}
\end{pspicture} &
\begin{pspicture}(0,0)(2,2)
\orientedEdge(1,1)(1,2)
\orientedEdge(2,1)(1,1)
\orientedEdge(1,0)(1,1)
\orientedEdge(1,1)(0,1)
\pscircle[fillstyle=solid](1,1){.6mm}
\end{pspicture} &
\begin{pspicture}(0,0)(2,2)
\orientedEdge(1,1)(1,2)
\orientedEdge(2,1)(1,1)
\orientedEdge(1,1)(1,0)
\orientedEdge(0,1)(1,1)
\pscircle[fillstyle=solid](1,1){.6mm}
\end{pspicture} &
\begin{pspicture}(0,0)(2,2)
\orientedEdge(1,2)(1,1)
\orientedEdge(1,1)(2,1)
\orientedEdge(1,0)(1,1)
\orientedEdge(1,1)(0,1)
\pscircle[fillstyle=solid](1,1){.6mm}
\end{pspicture} \\[1em]
$\alpha(\varphi | \eta)$ & $\alpha'(\varphi | \eta)$ & $\beta(\varphi | \eta)$ & $\beta'(\varphi | \eta)$ & $\gamma(\varphi | \eta)$ & $\gamma'(\varphi | \eta)$
\end{tabular}
\caption{}
\label{ThreeColoringToSixVertex}
\end{figure}
where $r$ is one of the colors $\overline 0$, $\overline 1$, and $\overline 2$.  The corresponding weights of the work~\cite{Str82} are expressed via standard elliptic theta functions of nome $p = \rme^{\rmi \pi \tau}$~\cite{WhiWat27} as follows
\begin{align}
&\alpha_r(\varphi | \lambda, p) = \alpha'_r(\varphi | \lambda, p) = \zeta_r^{1 / 4 + 3 \varphi / 4 \pi}(\lambda, p) \frac{\theta_1(\pi / 3 - \varphi | p)}{\theta_1(2 \pi / 3 | p)}, \label{e:1} \\
&\beta_r(\varphi | \lambda, p) = \beta'_r(\varphi | \lambda, p) = \zeta_r^{1 / 4 -  3 \varphi / 4 \pi}(\lambda, p)  \frac{\theta_1(\pi / 3 + \varphi | p)}{\theta_1(2 \pi / 3 | p)}, \\
&\gamma_r(\varphi | \lambda, p) = \frac{\zeta^{1 / 6 + \varphi / 2 \pi}_{r + \overline 1}(\lambda, p)}{\zeta^{1 / 6 + \varphi / 2 \pi}_r(\lambda, p)} \frac{\theta_4(\lambda + 2 \pi (r + 1 / 2) / 3 + \varphi | p)}{\theta_4(\lambda + 2 \pi r / 3 | p)}, \\
&\gamma'_r(\varphi | \lambda, p) = \frac{\zeta^{1 / 6 + \varphi / 2 \pi}_{r - \overline 1}(\lambda, p)}{\zeta^{1 / 6 + \varphi / 2 \pi}_r(\lambda, p)} \frac{\theta_4(\lambda + 2 \pi (r - 1 / 2) / 3 - \varphi | p)}{\theta_4(\lambda + 2 \pi r / 3 | p)}. \label{e:4}
\end{align}
Here $\varphi$ is the spectral parameter associated with the central vertex of a four-face configuration, $\lambda$ is an arbitrary fixed parameter, and
\begin{equation}
\zeta_r(\lambda, p) = \frac{\theta_4(\lambda + 2 \pi (r - 1) / 3 | p) \theta_4(\lambda +  2 \pi (r + 1) / 3 | p)}{\theta_4^2(\lambda + 2 \pi r / 3 | p)}. \label{e:5}
\end{equation}
It is useful to have in mind that\footnote{Below, when it does not lead to misunderstanding, we do not write up explicitly dependence on the fixed parameters.}
\begin{equation}
\zeta_{\overline 0} \, \zeta_{\overline 1} \, \zeta_{\overline 2} = 1, \label{e:6}
\end{equation}
and that
\begin{align}
&\alpha_{r + \overline 1}(\varphi | \lambda) = \alpha_r(\varphi | \lambda + 2 \pi / 3), &&\alpha'_{r + \overline 1}(\varphi | \lambda) = \alpha'_r(\varphi | \lambda + 2 \pi / 3), \label{e:7} \\
&\beta_{r + \overline 1}(\varphi | \lambda) = \alpha_r(\varphi | \lambda + 2 \pi / 3), &&\beta'_{r + \overline 1}(\varphi | \lambda) = \beta'_r(\varphi | \lambda + 2 \pi / 3), \\
&\gamma_{r + \overline 1}(\varphi | \lambda) = \gamma_r(\varphi | \lambda + 2 \pi / 3), &&\gamma'_{r + \overline 1}(\varphi | \lambda) = \gamma'_r(\varphi | \lambda + 2 \pi / 3). \label{e:8}
\end{align}

For $\varphi = 0$ we have
\begin{gather*}
\alpha_r(0) = \alpha'_r(0) = \zeta_r^{1/4},  \qquad \beta_r(0) = \beta'_r(0) =  \zeta_r^{1/4}, \\
\gamma_r(0) = \zeta_{r + \overline 1}^{1/2} \, \zeta_r^{1/2}, \qquad \gamma'_r(0) = \zeta_{r - \overline 1}^{1/2}  \, \zeta_r^{1/2},
\end{gather*}
and, using the equality (\ref{e:6}), it is not difficult to see that we come to the Baxter's three-coloring model with
\[
z_r = \zeta_r.
\]
It follows from the equality (\ref{e:6}) that $z_{\overline 0} \,z_{\overline 1} \, z_{\overline 2} = 1$, but this fact does not lead to any loss of generality.

In 1961 Lenard remarked that there is a correspondence between the three-co\-lo\-rings and the states of the six-vertex model~\cite{Len61}. Recall that the six-vertex model is defined on a square lattice. A state of the model is specified by a choice of the direction of each internal edge by placing an arrow on it. The arrows obey the rule, called the ice condition, that at every vertex there are exactly two arrows pointing in and two arrows pointing out. There are six possible configurations of arrows at each vertex, see the second row of pictures in Figure~\ref{ThreeColoringToSixVertex}, hence the name of the model.

To establish a correspondence between the three-colorings and the states of the six-vertex model,  take a three-coloring, remove the boundary edges of the lattice and place arrows on the internal edges in accordance with the following rule. Consider four edges containing a fixed four-valent vertex of the lattice and four faces containing these edges. The possible color combinations for such four-face sets are given by the first raw of the pictures in Figure~\ref{ThreeColoringToSixVertex}, where $r$ is one of the colors $\overline 0$, $\overline 1$, and $\overline 2$. Visit the selected faces moving anticlockwise. If intersecting an edge we see that the color changes by $+\overline 1$ we place on the edge a pointing in arrow, if the color changes by $-\overline 1$ we place a pointing out arrow. It is not difficult to get convinced that this rule is not contradictory, and we obtain the vertex configurations of the six-vertex model, see Figure~\ref{ThreeColoringToSixVertex}. The state of the six-vertex model, corresponding to the state of the three-coloring model given in Figure~\ref{ThreeColoringExample} is given in Figure~\ref{SixVertexExample}.
\begin{figure}[htb]
\centering
\begin{pspicture}(0,0)(6,5)
\orientedEdge(1,1)(1,0)
\orientedEdge(2,0)(2,1)
\orientedEdge(3,1)(3,0)
\orientedEdge(4,1)(4,0)
\orientedEdge(5,0)(5,1)
\orientedEdge(1,2)(1,1)
\orientedEdge(2,2)(2,1)
\orientedEdge(3,1)(3,2)
\orientedEdge(4,2)(4,1)
\orientedEdge(5,1)(5,2)
\orientedEdge(1,3)(1,2)
\orientedEdge(2,2)(2,3)
\orientedEdge(3,3)(3,2)
\orientedEdge(4,3)(4,2)
\orientedEdge(5,2)(5,3)
\orientedEdge(1,4)(1,3)
\orientedEdge(2,3)(2,4)
\orientedEdge(3,4)(3,3)
\orientedEdge(4,4)(4,3)
\orientedEdge(5,3)(5,4)
\orientedEdge(1,5)(1,4)
\orientedEdge(2,4)(2,5)
\orientedEdge(3,5)(3,4)
\orientedEdge(4,5)(4,4)
\orientedEdge(5,4)(5,5)
\orientedEdge(1,1)(0,1)
\orientedEdge(2,1)(1,1)
\orientedEdge(2,1)(3,1)
\orientedEdge(4,1)(3,1)
\orientedEdge(5,1)(4,1)
\orientedEdge(6,1)(5,1)
\orientedEdge(0,2)(1,2)
\orientedEdge(1,2)(2,2)
\orientedEdge(3,2)(2,2)
\orientedEdge(3,2)(4,2)
\orientedEdge(4,2)(5,2)
\orientedEdge(5,2)(6,2)
\orientedEdge(1,3)(0,3)
\orientedEdge(2,3)(1,3)
\orientedEdge(3,3)(2,3)
\orientedEdge(4,3)(3,3)
\orientedEdge(5,3)(4,3)
\orientedEdge(6,3)(5,3)
\orientedEdge(1,4)(0,4)
\orientedEdge(2,4)(1,4)
\orientedEdge(3,4)(2,4)
\orientedEdge(4,4)(3,4)
\orientedEdge(5,4)(4,4)
\orientedEdge(6,4)(5,4)
\multips(1,1)(0,1){4}{
  \multips(0,0)(1,0){5}{
    \pscircle[fillstyle=solid](0,0){.08}}}
\end{pspicture}
\caption{}
\label{SixVertexExample}
\end{figure}
It is also evident that the established correspondence is three-to-one, and if we fix the color of any face of the lattice it becomes one-to-one.

In the present paper we show that similarity between the three-coloring model and the six-vertex model extends further. Namely, choose the weights of the six-vertex model as
\begin{gather}
\alpha(\varphi | \eta) = \alpha'(\varphi | \eta) =  \frac{\sin(\eta / 2 - \varphi )}{\sin \eta}, \qquad \beta(\varphi | \eta) = \beta'(\varphi | \eta) = \frac{\sin(\eta / 2 + \varphi)}{\sin \eta}, \label{e:9a} \\
\gamma(\varphi | \eta) = \gamma'(\varphi | \eta) = 1, \label{e:9b}
\end{gather}
see Figure~\ref{ThreeColoringToSixVertex} for the correspondence between the functions and the vertex configurations. The parameter $\eta$ is called the crossing parameter. It is known that for $\eta = 2 \pi / 3$ and the domain wall boundary conditions the partition function of the inhomogeneous six-vertex model satisfies some simple functional equations \cite{Str05, RazStr04}, see Section~\ref{SVDW} for details. We show that the partition function of the three-coloring model for appropriate boundary conditions satisfies a similar equation.

In the papers~\cite{Str05, RazStr04} to obtain the functional equations for the six-vertex model the representation of the partition function via the Izergin--Korepin determinant~\cite{Ize87} was used. This representation is usually proved with the help of some recursion relations satisfied by the partition function which were found by Korepin~\cite{Kor82}. In Section~\ref{SVDW} we obtain the functional equations directly from the recursion relations. In Section~\ref{TCDW} we find recursion relations satisfied by the partition function of the three-coloring model for appropriate boundary conditions, which allow us to prove in this case functional equations similar to the functional equations of the six-vertex model.

One can show that the statistical three-coloring model is a partial case of the cyclic SOS (solid-on-solid) model \cite{PeaSea88} which in turn is a partial case of the eight-vertex SOS model \cite{Bax73}. Actually, the weights of the eight-vertex SOS model depends on a parameter $\eta$ analogous to the crossing parameter of the six-vertex model. The cyclic SOS model corresponds to the case when $\eta = 2 \pi / L$ for some positive integer $L$, and the statistical three-coloring model corresponds to the case when  $\eta = 2 \pi / 3$.

Recently, it was shown by Rosengren that the partition function of the eight-vertex SOS model can be represented as a linear combination of determinants \cite{Ros08}. In our case it is a linear combination of two determinants. One can use this representation to derive the functional equations, but we prefer to follow a direct way.

The weights used in the paper \cite{Ros08} for $\eta = 2 \pi / 3$ are different from the weights given by the formulas (\ref{e:1})--(\ref{e:4}). In the Appendix we show that these two sets of weights are connected by a change of the spectral parameter and a gauge transformation.

\section{Functional equations for six-vertex model from recursion relations}
\label{SVDW}

In this section we consider the six-vertex model with the domain wall boundary conditions~\cite{Kor82}. Here the boundary horizontal arrows point in and the boundary vertical arrows point out, see an example in Figure~\ref{DomainWallSixVertex}.
\begin{figure}[htb]
\centering
\begin{pspicture}(0,-.8)(5.8,5)
\orientedEdge(0,1)(1,1)
\orientedEdge(0,2)(1,2)
\orientedEdge(0,3)(1,3)
\orientedEdge(0,4)(1,4)
\orientedEdge(5,1)(4,1)
\orientedEdge(5,2)(4,2)
\orientedEdge(5,3)(4,3)
\orientedEdge(5,4)(4,4)
\orientedEdge(1,1)(1,0)
\orientedEdge(2,1)(2,0)
\orientedEdge(3,1)(3,0)
\orientedEdge(4,1)(4,0)
\orientedEdge(1,4)(1,5)
\orientedEdge(2,4)(2,5)
\orientedEdge(3,4)(3,5)
\orientedEdge(4,4)(4,5)
\multips(1,1)(0,1){4}{
\psline(0,0)(3,0)
}
\multips(1,1)(1,0){4}{
\psline(0,0)(0,3)
}
\multips(1,1)(0,1){4}{
  \multips(0,0)(1,0){4}{
    \pscircle[fillstyle=solid](0,0){.08}}}
\rput(1,-.4){$\psi_1$}
\rput(2,-.4){$\psi_2$}
\rput(3,-.4){$\psi_3$}
\rput(4,-.4){$\psi_4$}
\rput(5.5,4){$\chi_1$}
\rput(5.5,3){$\chi_2$}
\rput(5.5,2){$\chi_3$}
\rput(5.5,1){$\chi_4$}
\end{pspicture}
\caption{}
\label{DomainWallSixVertex}
\end{figure}
We label the horizontal lines by the variables $\chi_i$, $i = 1, \ldots, n$, and the vertical lines by the variables  $\psi_i$, $i = 1, \ldots, n$. The vertex at the intersection of the line labeled by $\chi_i$ and the line labeled by $\psi_j$ acquires the spectral parameter $\chi_i - \psi_j$. The weight of a state of the model is the product of the weights of the vertices, and the partition function $Z_n(\{\chi\}; \{\psi\})$ is the sum of the weights of all possible states.

The weights defined by the formulas (\ref{e:9a}) and (\ref{e:9b}) satisfy the Yang--Baxter equations. Using this fact, one can show that the partition function is separately symmetric in the variables $\chi_i$ and $\psi_i$, see, for example, \cite{KorBogIze93, Kup96}.

Define the function
\[
F_n(\{\chi\}; \{\psi\}) = \prod_{\substack{i,j = 1 \\ i < j}}^n \sin (\chi_i - \chi_j) \prod_{i,j = 1}^n \sin (\chi_i - \psi_j)  \prod_{\substack{i,j = 1 \\ i < j}}^n \sin (\psi_i - \psi_j) Z_n(\{\chi\}; \{\psi\}).
\]
Our goal is to prove that for $\eta = 2 \pi / 3$ the function $F_n(\{\chi\}; \{\psi\})$ satisfies the functional equations
\begin{equation}
\sum_{s =0}^2 F_n(\chi_1, \ldots, \chi_k + 2 \pi s / 3, \ldots, \chi_n; \{\psi\}) = 0 \label{e:10}
\end{equation}
and the functional equations\footnote{We write the equations (\ref{e:11}) with a minus sign before $2 \pi s / 3$ in the right hand side for similarity with the corresponding equations arising in the case of the three-coloring model. In the case of the six-vertex model we can use a plus sign as well.}
\begin{equation}
\sum_{s=0}^2 F_n(\{\chi\}; \psi_1, \ldots, \psi_k - 2 \pi s / 3, \ldots, \psi_n) = 0. \label{e:11}
\end{equation}
These equations can be used to solve some enumeration problems for the alternating-sign matrices. In particular, it is possible to reproduce \cite{Str05} the refined enumeration of the alternating-sign matrices conjectured by Mills, Robbins and Rumsey~\cite{MilRobRum83,MilRobRum82} and proved by Zeilberger~\cite{Zei96}.

We start with the case when $\eta$ is arbitrary. Consider the vertex on the intersection of the lines with the labels $\chi_n$ and $\psi_n$. There are two possible configurations of this vertex, the second and fourth ones in Figure~\ref{ThreeColoringToSixVertex}. They have the Boltzmann weights $\sin (\eta/2 - \chi_n + \psi_n)/\sin \eta$ and $1$ respectively. If we put $\chi_n = \psi_n + \eta / 2$, then only the lattice configurations corresponding to the second possibility will give a nonzero contribution to the partition function. Here the configuration of the vertices belonging to the bottom row and the left column become fixed, see Figure \ref{SixVertexRecursion} for an example.
\begin{figure}
\centering
\begin{pspicture}(0,-.8)(5.8,5)
\orientedEdge(0,1)(1,1)
\orientedEdge(0,2)(1,2)
\orientedEdge(0,3)(1,3)
\orientedEdge(0,4)(1,4)
\orientedEdge(5,1)(4,1)
\orientedEdge(5,2)(4,2)
\orientedEdge(5,3)(4,3)
\orientedEdge(5,4)(4,4)
\orientedEdge(1,1)(1,0)
\orientedEdge(2,1)(2,0)
\orientedEdge(3,1)(3,0)
\orientedEdge(4,1)(4,0)
\orientedEdge(1,4)(1,5)
\orientedEdge(2,4)(2,5)
\orientedEdge(3,4)(3,5)
\orientedEdge(4,4)(4,5)
\orientedEdge(1,1)(2,1)
\orientedEdge(2,1)(3,1)
\orientedEdge(3,1)(4,1)
\orientedEdge(4,1)(4,2)
\orientedEdge(4,2)(4,3)
\orientedEdge(4,3)(4,4)
\orientedEdge(1,2)(1,1)
\orientedEdge(2,2)(2,1)
\orientedEdge(3,2)(3,1)
\orientedEdge(4,2)(3,2)
\orientedEdge(4,3)(3,3)
\orientedEdge(4,4)(3,4)
\multips(1,2)(0,1){3}{
\psline(0,0)(2,0)
}
\multips(1,2)(1,0){3}{
\psline(0,0)(0,2)
}
\multips(1,1)(0,1){4}{
  \multips(0,0)(1,0){4}{
    \pscircle[fillstyle=solid](0,0){.08}}}
\rput(1,-.4){$\psi_1$}
\rput(2,-.4){$\psi_2$}
\rput(3,-.4){$\psi_3$}
\rput(4,-.4){$\psi_4$}
\rput(5.5,4){$\chi_1$}
\rput(5.5,3){$\chi_2$}
\rput(5.5,2){$\chi_3$}
\rput(5.5,1){$\chi_4$}
\end{pspicture}
\caption{}
\label{SixVertexRecursion}
\end{figure}
The configurations of the remaining vertices correspond to the configurations of the $(n-1) \times (n-1)$ lattice with the domain wall boundary conditions. Collecting the weights of the fixed vertices, we come to the following recursion relation
\begin{multline*}
\left. Z_n(\chi_1, \ldots, \chi_{n-1}, \chi_n; \psi_1, \ldots, \psi_{n-1}, \psi_n) \right|_{\chi_n = \psi_n + \eta / 2} \\
= \sin^{2-2n} \eta \prod_{i = 1}^{n-1} \sin (\chi_i - \psi_n + \eta / 2) \prod_{i = 1}^{n-1} \sin (\psi_n - \psi_i + \eta) \\
\times Z_{n-1}(\chi_1, \ldots, \chi_{n-1}; \psi_1, \ldots, \psi_{n-1}).
\end{multline*}
Actually, since the function $Z_n(\{\chi\}; \{\psi\})$ is symmetric in the variables $\chi_i$ and $\psi_i$, we have $n^2$ recursion relations
\begin{multline}
\left. Z_n(\chi_1, \ldots, \chi_k, \ldots, \chi_n; \psi_1, \ldots, \psi_\ell, \ldots, \psi_n) \right|_{\chi_k = \psi_\ell + \eta / 2} \\
= \sin^{2-2n} \eta \prod_{\substack{i = 1 \\ i \ne k}}^n \sin (\chi_i - \psi_\ell + \eta / 2) \prod_{\substack{i = 1 \\ i \ne \ell}}^n \sin ( \psi_\ell - \psi_i + \eta) \\
\times Z_{n-1}(\chi_1, \ldots, \widehat{\chi_k}, \ldots, \chi_n; \psi_1, \ldots, \widehat{\psi_\ell}, \ldots, \psi_n). \label{e:12}
\end{multline}

Now we start with the vertex on the intersection of the lines with the labels $\chi_1$ and $\psi_n$. In a similar way as above we obtain $n^2$ recursion relations
\begin{multline}
\left. Z_n(\chi_1, \ldots, \chi_k, \ldots, \psi_n; \psi_1, \ldots, \psi_\ell, \ldots, \psi_n) \right|_{\chi_k = \psi_\ell - \eta/2} \\
= \sin^{2-2n} \eta \prod_{\substack{i = 1 \\ i \ne k}}^n \sin (\chi_i - \psi_\ell  - \eta / 2) \prod_{\substack{i = 1 \\ i \ne \ell}}^n \sin (\psi_\ell - \psi_i - \eta) \\
\times Z_{n-1}(\chi_1, \ldots, \widehat{\chi_k}, \ldots, \chi_n; \psi_1, \ldots, \widehat{\psi_\ell}, \ldots, \psi_n). \label{e:13}
\end{multline}

The recursion relations (\ref{e:12}) and (\ref{e:13}) are valid for an arbitrary $\eta$. Now assume that $\eta = 2 \pi / 3$. Using the identity
\[
\sin(\varphi) \sin(\varphi + \pi / 3) \sin(\varphi + 2 \pi /3) = \frac{1}{4} \sin (3 \varphi)
\]
and the recursion relations (\ref{e:12}) and (\ref{e:13}), we can prove that
\begin{multline}
\left. F_n(\chi_1, \ldots, \chi_k, \ldots, \chi_n; \psi_1, \ldots, \psi_\ell, \ldots, \psi_n) \right|_{\chi_k = \psi_\ell \pm \pi/3} \\
= \pm  (-1)^{n-1} 4^{2 - 2 n} \sin^{3 - 2 n} (2 \pi / 3) \prod_{\substack{i = 1 \\ i \ne k}}^n \sin [3 (\psi_\ell - \chi_i)] \prod_{\substack{i = 1 \\ i \ne \ell}}^n \sin [3 (\psi_\ell - \psi_i)] \\
\times F_{n-1}(\chi_1, \ldots, \widehat{\chi_k}, \ldots, \chi_n; \psi_1, \ldots, \widehat{\psi_\ell}, \ldots, \psi_n).
\label{e:14}
\end{multline}

Define the functions
\[
S_{n, k}(\chi_1, \ldots, \chi_k, \ldots, \chi_n; \{\psi\}) = \sum_{s=0}^2 F_n(\chi_1, \ldots, \chi_k + 2 \pi s / 3, \ldots, \chi_n; \{\psi\}),
\]
and prove by induction that $S_{n, k}(\{\chi\}; \{\psi\}) = 0$. Actually, due to the skew-symmetry of the functions $F_n(\{\chi\}; \{\psi\})$ in the variables $\chi_i$, it suffices to prove that
\[
S_{n, 1}(\{\chi\}; \{\psi\}) = 0.
\]
It is not difficult to see that
\[
S_{1,1}(\chi_1; \psi_1) = \sum_{r=0}^2 \sin(\chi_1 - \psi_1 + 2 \pi / 3) = 0.
\]
Assume now that $S_{n-1, 1}(\{\chi\}; \{\psi\}) = 0$ for some $n > 1$. The recursion relations (\ref{e:14}) give
\begin{multline*}
\left. S_{n, 1}(\chi_1, \ldots, \chi_{n-1}, \chi_n; \psi_1, \ldots, \psi_\ell, \ldots, \psi_n) \right|_{\chi_n = \psi_\ell \pm \pi/3} \\
= \mp  (-1)^n 4^{2 - 2 n} \sin^{3 - 2 n} (2 \pi / 3) \prod_{i = 1 }^{n-1} \sin [3 (\psi_\ell - \chi_i)] \prod_{\substack{i = 1 \\ i \ne \ell}}^n \sin [3 (\psi_\ell - \psi_i)] \\
\times S_{n-1, 1}(\chi_1, \ldots, \chi_{n-1}; \psi_1, \ldots, \widehat{\psi_\ell}, \ldots, \psi_n) = 0.
\end{multline*}

It is clear that for the domain wall boundary condition any row or a column of a state has an odd number of $\gamma$-type and $\gamma'$-type vertices. Therefore, we have, in particular, that
\begin{equation}
Z_n(\chi_1, \ldots, \chi_n + \pi; \{\psi\}) = (-1)^{n-1} Z_n(\chi_1, \ldots, \chi_n; \{\psi\}).  \label{e:15}
\end{equation}
With respect to the variable $\chi_n$ the partition function $Z_n(\{\chi\}; \{\psi\})$ is a trigonometric polynomial of order less or equal to $n-1$, and  $S_{n, 1}(\{\chi\}; \{\psi\})$ is a trigonometric polynomial of order less or equal to $3 n -2$.  As we showed just above, the function $S_{n, 1}(\{\chi\}; \{\psi\})$ has zeros at the points $\chi_n = \psi_\ell \pm \pi / 3$, $\ell = 1, \ldots, n$. Moreover, by construction it also has zeros at the points $\chi_n = \chi_\ell$, $\ell = 2, \ldots, n-1$ and at the points $\chi_n = \psi_\ell$, $\ell = 1, \ldots, n$. The relation (\ref{e:15}) doubles all these zeros. Hence, we have $8 n - 4$ zeros in the interval $0 \le \chi_n < 2 \pi$. This is possible only if $S_{n, 1}(\{\chi\}; \{\psi\}) = 0$. Thus, we proved the equations (\ref{e:10}). The equations (\ref{e:11}) can be proved in the same way.

\section{Functional equations for three-coloring model with domain wall boundary conditions}
\label{TCDW}

For the three-coloring model there is a natural analogue of the domain wall boundary conditions of the six-vertex model.  Here if one starts with any boundary face and walks anticlockwise along the boundary then the color changes by $+ \overline 1$ from face to face for the vertical boundaries, and by $- \overline 1$ for the horizontal boundaries.  An example is given in Figure \ref{DomainWallThreeColoring}.
\begin{figure}[htb]
\centering
\begin{pspicture}(0,-.8)(5.8,5.2)
\multips(0,0)(0,1){6}{
  \psline(0,0)(5,0)}
\multips(0,0)(1,0){6}{
  \psline(0,0)(0,5)  }
\multips(0,0)(0,1){6}{
  \multips(0,0)(1,0){6}{
    \pscircle[fillstyle=solid](0,0){.08}}}
\rput(.5,4.5){$\scriptstyle \overline 2$}
\rput(1.5,4.5){$\scriptstyle \overline 0$}
\rput(2.5,4.5){$\scriptstyle \overline 1$}
\rput(3.5,4.5){$\scriptstyle \overline 2$}
\rput(4.5,4.5){$\scriptstyle \overline 0$}
\rput(4.5,3.5){$\scriptstyle \overline 2$}
\rput(4.5,2.5){$\scriptstyle \overline 1$}
\rput(4.5,1.5){$\scriptstyle \overline 0$}
\rput(4.5,0.5){$\scriptstyle \overline 2$}
\rput(.5,3.5){$\scriptstyle \overline 0$}
\rput(.5,2.5){$\scriptstyle \overline 1$}
\rput(.5,1.5){$\scriptstyle \overline 2$}
\rput(.5,.5){$\scriptstyle \overline 0$}
\rput(1.5,.5){$\scriptstyle \overline 2$}
\rput(2.5,.5){$\scriptstyle \overline 1$}
\rput(3.5,.5){$\scriptstyle \overline 0$}
\rput(1,-.4){$\psi_1$}
\rput(2,-.4){$\psi_2$}
\rput(3,-.4){$\psi_3$}
\rput(4,-.4){$\psi_4$}
\rput(5.5,4){$\chi_1$}
\rput(5.5,3){$\chi_2$}
\rput(5.5,2){$\chi_3$}
\rput(5.5,1){$\chi_4$}
\end{pspicture}
\caption{}
\label{DomainWallThreeColoring}
\end{figure}
It is clear that in this case the number of face rows should coincide with the number of face columns. One can easily get convinced that the corresponding states of the six-vertex model satisfy the domain wall boundary conditions.

We will consider the inhomogeneous case when the internal horizontal lines are labeled by the variables $\chi_i$, $i = 1, \ldots, n$, and the internal vertical lines are labeled by the variables  $\psi_i$, $i = 1, \ldots, n$, see for example Figure~\ref{DomainWallThreeColoring}. Here and henceforth $n$ means the number of internal vertices in a row or in a column. With the vertex at the intersection of the line labeled by $\chi_i$ and the line labeled by $\psi_j$ we associate the spectral parameter $\chi_i - \psi_j$. We represent the total partition function $Z_n(\{\chi\}; \{\psi\})$ as the sum of partial partition functions:
\[
Z_n(\{\chi\}; \{\psi\}) = \sum_{r \in \bbZ_3} Z^r_n(\{\chi\}; \{\psi\}),
\]
where  $r$ is the color of the left topmost face of the lattice. Note that the equalities (\ref{e:7})--(\ref{e:8}) imply that
\[
Z^{r + \overline 1}_n(\{\chi\}; \{\psi\} | \lambda, p) = Z^r_n(\{\chi\}; \{\psi\} | \lambda + 2 \pi / 3, p).
\]

We denote the weight corresponding to the four-face configuration
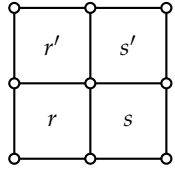
\[
\begin{pspicture}(-.1,-.1)(2.1,2.1)
\psline (0,0)(2,0)
\psline (0,1)(2,1)
\psline (0,2)(2,2)
\psline (0,0)(0,2)
\psline (1,0)(1,2)
\psline (2,0)(2,2)
\multips(0,0)(0,1){3}{
  \multips(0,0)(1,0){3}{
    \pscircle[fillstyle=solid](0,0){.08}}}
\rput(.5,1.5){$\scriptstyle r'$}
\rput(1.5,1.5){$\scriptstyle s'$}
\rput(1.5,.5){$\scriptstyle s$}
\rput(.5,.5){$\scriptstyle r$}
\end{pspicture}
\]
by $W_{r \, s}^{r' s'}(\varphi)$. As we noted above, these weights satisfy the Yang--Baxter equations which has the explicit form
\begin{multline}
\sum_{t \in \bbZ_3} W_{r' t}^{r'' s''}(\varphi) W_{r \, s}^{r' t}(\varphi') W_{t \, s}^{s'' s'}(\varphi - \varphi' - \pi / 3) \\
= \sum_{t \in \bbZ_3} W_{r' r}^{r'' t}(\varphi - \varphi' - \pi / 3) W_{t \, s'}^{r'' s''}(\varphi') W_{r\,  s}^{t \, s'}(\varphi)
\label{e:16}
\end{multline}
and guarantees the integrability of the model.

It is not difficult to get convinced that if weights $W^{r' s'}_{r \, s}(\varphi)$ satisfy the Yang--Baxter equations (\ref{e:16}), then for arbitrary non-zero constants $C_r$ gauge transformed weights $\widetilde W^{r' s'}_{r s}(\varphi)$, defined as
\begin{equation}
\widetilde W^{r' s'}_{r \, s}(\varphi) =\frac{C_r^{}}{C_{s'}} \, \frac{\Phi_{r'}(\varphi) \Phi_s(\varphi)}{\Phi_r(\varphi) \Phi_{s'}(\varphi)} \,  W^{r' s'}_{r \, s}(\varphi) \label{e:17}
\end{equation}
also satisfy the Yang--Baxter equations relation provided that
\[
\Phi_r(\varphi - \varphi' - \pi / 3) = \frac{\Phi_r(\varphi)}{\Phi_r(\varphi')}.
\]
The corresponding partition sums coincide in the case of the toroidal boundary conditions, and may differ by a factor in other cases.

We apply the gauge transformation (\ref{e:17}) with
\[
C_r = 1, \qquad \Phi_r(\varphi) = \zeta_r^{1 / 12 + \varphi / 4 \pi}
\]
to the weights of the three-coloring model defined by the functions (\ref{e:1})--(\ref{e:4}). As the result we come to the weights defined by the functions
\begin{align}
&\tilde \alpha_r(\varphi) = \tilde \alpha'_r(\varphi) = \frac{\theta_1(\pi / 3 - \varphi)}{\theta_1(2 \pi / 3)}, \label{e:18} \\
&\tilde  \beta_r(\varphi) = \tilde  \beta'_r(\varphi) = \zeta_r^{1/2} \,  \frac{\theta_1(\pi / 3 + \varphi)}{\theta_1(2 \pi / 3)}, \\
&\tilde \gamma_r(\varphi) = \frac{\theta_4(\lambda + 2 \pi (r + 1 / 2) / 3 + \varphi)}{\theta_4(\lambda + 2 \pi r  / 3)}, \\
& \tilde  \gamma{}'_r(\varphi) = \frac{\theta_4(\lambda + 2 \pi (r - 1 / 2) / 3 - \varphi)}{\theta_4(\lambda + 2 \pi r / 3)}. \label{e:21}
\end{align}
The corresponding partial partition functions for our boundary conditions are connected by the equality
\[
\widetilde Z{}^r_n(\{\chi\}; \{\psi\}) = \prod_{i=1}^n \left[ \frac{\Phi_{r + i - \overline 1}(\chi_i - \psi_i + \chi_{n-i+1} - \psi_{n-i+1})}{\Phi_{r + i}(\chi_i - \psi_i + \chi_{n-i+1} - \psi_{n-i+1})} \right] Z^r_n(\{\chi\}; \{\psi\}).
\]
Using a trick similar to that used for the six-vertex model \cite{KorBogIze93, Kup96}, one can show that the Yang--Baxter equations lead to the separate symmetry of $\widetilde Z^r_n(\{\chi\}; \{\psi\})$ in the variables $\chi_i$ and $\psi_i$.

Consider the vertex on the intersection of the line with the label $\chi_n$ and the line with the label $\psi_n$. There are two possible four-face configurations for this vertex, the second and fifth ones in Figure~\ref{ThreeColoringToSixVertex}. They have the Boltzmann weights $\tilde \alpha'_r (\chi_n - \psi_n)$ and $\tilde \gamma_r(\chi_n - \psi_n)$ respectively. If we put $\chi_n = \psi_n + \pi / 3$, then only the configurations corresponding to the second possibility will give a nonzero contribution to the partition function. Here the four-face configurations for the vertices belonging to the internal bottom row and the internal left column become fixed, see Figure \ref{ThreeColoringRecursion} for an example.
\begin{figure}[htb]
\centering
\begin{pspicture}(0,-.8)(5.8,5.2)
\multips(0,0)(0,1){6}{
  \psline(0,0)(5,0)}
\multips(0,0)(1,0){6}{
  \psline(0,0)(0,5)  }
\multips(0,0)(0,1){6}{
  \multips(0,0)(1,0){6}{
    \pscircle[fillstyle=solid](0,0){.08}}}
\rput(.5,4.5){$\scriptstyle \overline 2$}
\rput(1.5,4.5){$\scriptstyle \overline 0$}
\rput(2.5,4.5){$\scriptstyle \overline 1$}
\rput(3.5,4.5){$\scriptstyle \overline 2$}
\rput(4.5,4.5){$\scriptstyle \overline 0$}
\rput(4.5,3.5){$\scriptstyle \overline 2$}
\rput(4.5,2.5){$\scriptstyle \overline 1$}
\rput(4.5,1.5){$\scriptstyle \overline 0$}
\rput(4.5,0.5){$\scriptstyle \overline 2$}
\rput(.5,3.5){$\scriptstyle \overline 0$}
\rput(.5,2.5){$\scriptstyle \overline 1$}
\rput(.5,1.5){$\scriptstyle \overline 2$}
\rput(.5,.5){$\scriptstyle \overline 0$}
\rput(1.5,.5){$\scriptstyle \overline 2$}
\rput(2.5,.5){$\scriptstyle \overline 1$}
\rput(3.5,.5){$\scriptstyle \overline 0$}
\rput(1.5,1.5){$\scriptstyle \overline 1$}
\rput(2.5,1.5){$\scriptstyle \overline 0$}
\rput(3.5,1.5){$\scriptstyle \overline 2$}
\rput(3.5,2.5){$\scriptstyle \overline 0$}
\rput(3.5,3.5){$\scriptstyle \overline 1$}
\rput(1,-.4){$\psi_1$}
\rput(2,-.4){$\psi_2$}
\rput(3,-.4){$\psi_3$}
\rput(4,-.4){$\psi_4$}
\rput(5.5,4){$\chi_1$}
\rput(5.5,3){$\chi_2$}
\rput(5.5,2){$\chi_3$}
\rput(5.5,1){$\chi_4$}
\end{pspicture}
\caption{}
\label{ThreeColoringRecursion}
\end{figure}
The configurations of the remaining faces correspond to the configuration of the \hbox{$(n-1) \times (n-1)$} lattice with the domain wall boundary. Therefore, we have the following recursion relation
\begin{multline*}
\left. \widetilde Z^r_n(\chi_1, \ldots, \chi_{n-1}, \chi_n; \psi_1, \ldots, \psi_{n-1}, \psi_n) \right|_{\chi_n = \psi_n + \pi / 3} \\
= \tilde \gamma_r(2 \pi / 3) \prod_{i = 1}^{n-1} \tilde \beta'_{r + n - i} (\chi_i - \psi_n) \prod_{i = 1}^{n-1} \tilde \beta_{r + n - i} (\psi_n - \psi_i + \pi / 3) \\
\times \widetilde Z^r_{n-1}(\chi_1, \ldots, \chi_{n-1}; \psi_1, \ldots, \psi_{n-1}).
\end{multline*}
Using the definitions (\ref{e:18})--(\ref{e:21}) and (\ref{e:5}), we rewrite this relation as
\begin{multline*}
\left. \widetilde Z^r_n(\chi_1, \ldots, \chi_{n-1}, \chi_n; \psi_1, \ldots, \psi_{n-1}, \psi_n) \right|_{\chi_n = \psi_n + \pi / 3} =  \theta_1^{2 - 2n}(2 \pi / 3) \\
\times \frac{\theta_4(\lambda + 2 \pi (r + n) / 3)} {\theta_4(\lambda + 2 \pi (r + n - 1) / 3)}  \prod_{i = 1}^{n-1} \theta_1(\chi_i - \psi_n  + \pi / 3) \prod_{i = 1}^{n-1} \theta_1(\psi_n - \psi_i + 2 \pi / 3) \\
\times \widetilde Z^r_{n-1}(\chi_1, \ldots, \chi_{n-1}; \psi_1, \ldots, \psi_{n-1}).
\end{multline*}
Since the functions $\widetilde Z^r_n(\{\chi\}; \{\psi\})$ are symmetric in the variables $\chi_i$ and $\psi_i$, we have actually $n^2$ recursion relations:
\begin{multline}
\left. \widetilde Z^r_n(\chi_1, \ldots, \chi_k, \ldots, \chi_n; \psi_1, \ldots, \psi_\ell,  \ldots, \psi_n) \right|_{\chi_k = \psi_\ell + \pi / 3} = \theta_1^{2 - 2n}(2 \pi / 3) \\
\times \frac{\theta_4(\lambda + 2 \pi (r + n) / 3)} {\theta_4(\lambda + 2 \pi (r + n - 1) / 3)}   \prod_{\substack{i = 1 \\ i \ne k}}^n \theta_1(\chi_i - \psi_\ell + \pi / 3) \prod_{\substack{i = 1 \\ i \ne \ell}}^n \theta_1(\psi_\ell - \psi_i + 2 \pi / 3) \\
\times \widetilde Z^r_{n-1}(\chi_1, \ldots, \widehat \chi_k, \ldots, \chi_n; \psi_1, \ldots, \widehat \psi_\ell, \ldots, \psi_n). \label{e:22}
\end{multline}

Starting with the right topmost face and exploring again the symmetry of the functions $\widetilde Z^r_n(\chi; \psi)$ in the variables $\chi_i$ and $\psi_i$, we obtain another set of recursion relations:
\begin{multline}
\left. \widetilde Z^r_n(\chi_1, \ldots, \chi_k, \ldots, \chi_n; \psi_1, \ldots, \psi_\ell,  \ldots, \psi_n) \right|_{\chi_k = \psi_\ell - \pi / 3} = \theta_1^{2 - 2n}(2 \pi / 3) \\
\times \prod_{\substack{i = 1 \\ i \ne k}}^n \theta_1(\chi_i - \psi_\ell - \pi / 3) \prod_{\substack{i = 1 \\ i \ne \ell}}^n \theta_1(\psi_\ell - \psi_i - 2 \pi / 3) \\
\times \widetilde Z^{r + \overline 1}_{n-1}(\chi_1, \ldots, \widehat \chi_k, \ldots, \chi_n; \psi_1, \ldots, \widehat \psi_\ell, \ldots, \psi_n). \label{e:23}
\end{multline}

The relations (\ref{e:22}) and (\ref{e:23}) are similar to the relations (\ref{e:12}) and (\ref{e:13}) of the preceding Section. Now we will find the analogue of the equations (\ref{e:10}) and (\ref{e:11}). To this end we start with introducing the functions
\begin{multline}
F^r_n(\{\chi\};  \{\psi\}) = \frac{1}{\theta_4(\lambda + 2 \pi (r + n) / 3)}  \\
\times \prod_{\substack{i,j = 1 \\ i < j}}^n \theta_1 (\chi_i - \chi_j) \prod_{i,j = 1}^n \theta_1(\chi_i - \psi_j)  \prod_{\substack{i,j = 1 \\ i < j}}^n \theta_1 (\psi_i - \psi_j)  \widetilde Z^r_n(\{\chi\}; \{\psi\}). \label{e:24}
\end{multline}
It is not difficult to get convinced that
\begin{equation}
\theta_1(\varphi | p) \theta_1(\varphi + \pi / 3 | p) \theta_1(\varphi + 2 \pi / 3 | p) = D(p) \theta_1(3 \varphi | p^3), \label{e:25}
\end{equation}
where $D(p)$ is given by the relation
\[
D(p) = \frac{\theta_1'(0 | p) \theta_1(\pi / 3 | p) \theta_1(2 \pi / 3 | p)}{3 \theta_1'(0 | p^3)}.
\]
Using the equality (\ref{e:25}) and the recursion relations (\ref{e:22}), we see that
\begin{multline}
\left. F^r_n(\chi_1, \ldots, \chi_k, \ldots, \chi_n; \psi_1, \ldots, \psi_\ell,  \ldots, \psi_n | \lambda, p) \right|_{\chi_k = \psi_\ell + \pi / 3} \\
= (-1)^{n-1} D^{2n - 2}(p) \theta_1^{3 - 2n}(2 \pi / 3 | p) \prod_{\substack{i = 1 \\ i \ne k}}^n \theta_1(3( \psi_\ell - \chi_i) | p^3) \prod_{\substack{i = 1 \\ i \ne \ell}}^n \theta_1(3(\psi_\ell - \psi_i) | p^3) \\
\times F^r_{n-1}(\chi_1, \ldots, \widehat \chi_k, \ldots, \chi_n; \psi_1, \ldots, \widehat \psi_\ell, \ldots, \psi_n | \lambda, p).
\label{e:26}
 \end{multline}
In a similar way the recursion relation (\ref{e:23}) gives
\begin{multline}
\left. F^r_n(\chi_1, \ldots, \chi_k, \ldots, \chi_n; \psi_1, \ldots, \psi_\ell,  \ldots, \psi_n | \lambda, p) \right|_{\chi_k = \psi_\ell - \pi / 3} \\
= - (-1)^{n-1} D^{2n - 2}(p) \theta_1^{3 - 2n}(2 \pi / 3 | p) \prod_{\substack{i = 1 \\ i \ne k}}^n \theta_1(3( \psi_\ell - \chi_i) | p^3) \prod_{\substack{i = 1 \\ i \ne \ell}}^n \theta_1(3(\psi_\ell - \psi_i) | p^3) \\
\times F^{r + \overline 1}_{n-1}(\chi_1, \ldots, \widehat \chi_k, \ldots, \chi_n; \psi_1, \ldots, \widehat \psi_\ell, \ldots, \psi_n | \lambda, p).
\label{e:27}
 \end{multline}

Now we define the functions
\[
S_{n,k}^r(\chi_1, \ldots, \chi_k, \ldots, \chi_n; \{\psi\}) = \sum_{s=0}^2 F_n^{r + s}(\chi_1, \ldots, \chi_k + 2 \pi s / 3, \ldots, \chi_n; \{\psi\}).
\]
It appears that $S_{n, k}^r(\{\chi\}, \{\psi\}) = 0$. To prove this fact we note first that it suffices to prove that $S_{n, 1}^{\overline 0}(\{\chi\}, \{\psi\}) = 0$. All other cases can be reduced to this one using the skew-symmetry of the functions $F^r_n(\{\chi\}, \{\psi\})$ in the variables $\chi_i$ and the equality
\[
F^{r + \overline 1}_n(\{\chi\}, \{\psi\} | \lambda, p) = F^r_n(\{\chi\}, \{\psi\} | \lambda + 2 \pi / 3, p),
\]
which follows from the relation
\[
\widetilde W_{r + \overline 1, r' + \overline 1}^{s + \overline 1, s' + \overline 1}(\varphi | \lambda, p) =\widetilde W_{r r'}^{s s'}(\varphi | \lambda + 2 \pi / 3, p).
\]

We prove the equality  $S_{n, 1}^{\overline 0}(\{\chi\}, \{\psi\}) = 0$ by induction on $n$. It is easy to see that
\[
F^r_1(\chi_1; \psi_1) =  \frac{\theta_1(\chi_1 - \psi_1) \theta_4(\lambda + \chi_1 - \psi_1 + 2 \pi (r + 1 / 2) / 3)}{\theta_4(\lambda + 2 \pi (r + 1) / 3) \theta_4(\lambda + 2 \pi r / 3)}.
\]
Hence, the function $S_{1, 1}^{\overline 0}(\chi_1; \psi_1)$ considered as a function on $\varphi = \chi_1 - \psi_1$ has the form
\begin{multline}
S_{1, 1}^{\overline 0}(\varphi) =  \frac{\theta_1(\varphi) \theta_4(\lambda + \varphi + \pi / 3)}{\theta_4(\lambda + 2 \pi / 3) \theta_4(\lambda)} \\
+ \frac{\theta_1(\varphi + 2 \pi / 3) \theta_4(\lambda + \varphi + 5 \pi / 3)}{\theta_4(\lambda + 4 \pi / 3) \theta_4(\lambda + 2 \pi /3)} + \frac{\theta_1(\varphi + 4 \pi / 3) \theta_4(\lambda + \varphi + 3 \pi)}{\theta_4(\lambda + 2 \pi) \theta_4(\lambda + 4 \pi / 3)}. \label{e:28}
\end{multline}
All terms of the sum in the right hand side of the equality (\ref{e:28}) have the same periodicity factor associated with the periods $\pi$ and $\pi \tau$. Dividing this sum by its first term, we obtain a doubly periodic function. It is not difficult to verify that this function has no poles, hence, it is a constant At $\varphi = - \pi / 3$ the function $S_{1, 1}^{\overline 0}(\varphi)$ turns to zero, but the first term of the sum in the right hand side of the equality (\ref{e:28}) is not zero at this point. Thus, the constant in question is zero, and $S_{1, 1}^{\overline 0}(\chi_1; \psi_1) = 0$. As we noted above, this means that $S_{1, 1}^r (\chi_1; \psi_1) = 0$ for any $r$.

Now we assume that $S^r_{n-1, 1}(\{\chi\}, \{\psi\}) = 0$ for some $n > 1$. It follows from the recursion relations (\ref{e:26}) that
\begin{multline*}
\left. S^r_{n, 1}(\chi_1, \ldots, \chi_{n-1}, \chi_n; \psi_1, \ldots, \psi_\ell,  \ldots, \psi_n | \lambda, p) \right|_{\chi_n = \psi_\ell + \pi / 3} \\
= (-1)^{n-1} D^{2n - 2}(p) \theta_1^{3 - 2n}(2 \pi / 3 | p) \prod_{i = 1}^{n-1} \theta_1(3(\psi_\ell - \chi_i) | p^3) \prod_{\substack{i = 1 \\ i \ne \ell}}^n \theta_1(3(\psi_\ell - \psi_i) | p^3) \\
\times S^r_{n - 1, 1}(\chi_1, \ldots, \chi_{n-1}; \psi_1, \ldots, \widehat \psi_\ell, \ldots, \psi_n | \lambda, p) = 0,
 \end{multline*}
and the recursion relations (\ref{e:27}) give
\begin{multline*}
\left. S^r_{n, 1}(\chi_1, \ldots, \chi_{n-1}, \chi_n; \psi_1, \ldots, \psi_\ell,  \ldots, \psi_n | \lambda, p) \right|_{\chi_n = \psi_\ell - \pi / 3} \\
= - (-1)^{n-1} D^{2n - 2}(p) \theta_1^{3 - 2n}(2 \pi / 3 | p) \prod_{i = 1}^{n-1} \theta_1(3(\psi_\ell - \chi_i) | p^3) \prod_{\substack{i = 1 \\ i \ne \ell}}^n \theta_1(3(\psi_\ell - \psi_i) | p^3) \\
\times S^{r + \overline 1}_{n - 1, 1}(\chi_1, \ldots, \chi_{n-1}; \psi_1, \ldots, \widehat \psi_\ell, \ldots, \psi_n | \lambda, p) = 0.
 \end{multline*}

Recall that for any state of the six-vertex model with the domain wall boundary conditions the number of $\gamma$-type and $\gamma'$-type vertices in any column and in any row of vertices is odd. It follows from the correspondence between the states of the three-coloring model and six-vertex model, described in Section \ref{s:1}, that for any state of the three-coloring model with the domain wall boundary conditions the total number of $\gamma$-type and $\gamma'$-type vertices in any column and in any row of the internal vertices is odd as well. Having in mind the equalities
\[
\theta_1(\varphi \pm \pi | p)=- \theta_1(\varphi | p), \qquad \theta_4(\varphi \pm \pi | p) = \theta_4(\varphi | p),
\]
we conclude that any summand of the partition function $\widetilde Z^r_n(\{\chi\}; \{\psi\})$ have the same periodicity factor associated with the period $\pi$ with respect to any variable $\chi_i$. The same is true for the summands of the functions $F^r_n(\{\chi\}; \{\psi\})$ and $S^r_{n, k}(\{\chi\}; \{\psi\})$.

Using the equalities
\[
\theta_1(\varphi + \pi \tau | p) = - p^{-1} \exp^{-2 \rmi \varphi} \theta_1(\varphi | p), \qquad \theta_4(\varphi + \pi \tau | p) = - p^{-1} \exp^{-2 \rmi \varphi} \theta_4(\varphi | p),
\]
we see that
\[
\widetilde W^{r' \, s'}_{r \, s}(\varphi + \pi \tau) = - p^{-1} \rme^{- 2 \rmi \varphi} \frac{\Psi_{r'} \, \Psi _s}{\Psi_r \Psi _{s'} } \widetilde W^{r' \, s'}_{r \, s}(\varphi),
\]
where
\[
\Psi_r = \rme^{\rmi (r - 1)[\pi (r + 1) / 3 + \lambda] \tau}.
\]
Now it is possible to show that any summand of the partition function $\widetilde Z^r_n(\{\chi\}; \{\psi\})$ have the same periodicity factor associated with the period $\pi \tau$ with respect to any variable $\chi_i$. The same is true for the summands of the functions $F^r_n(\{\chi\}; \{\psi\})$ and $S^r_{n, k}(\{\chi\}; \{\psi\})$.

Any summand of the function  $S^r_{n, 1}(\{\chi\}; \{\psi\})$ contains a product of $3n - 1$ theta functions whose arguments contain the variable $\chi_n$. Since all summands have the same periodicity factors, if we divide $S^r_{n, 1}(\{\chi\}; \{\psi\})$ by any of its summands we will obtain a doubly periodic function of $\chi_n$. Every elliptic theta function has  one and only one zero in a cell with the corners $\xi, \xi + \pi, \xi + \pi + \pi \tau, \xi + \tau$. Therefore, the order of the function obtained after division of the function $S^r_{n, 1}(\{\chi\}; \{\psi\})$ by any of its summands, is less or equal to $3 n - 1$. From the other hand, as we showed above the function $S^r_{n, 1}(\{\chi\}; \{\psi\})$ is zero for $\chi_n = \psi_\ell \pm 2 \pi / 3$, $\ell = 1, \ldots, n$. Moreover, it follows from the definition of the functions $S^r_{n, k}(\{\chi\}; \{\psi\})$ that $S^r_{n, 1}(\{\chi\}; \{\psi\})$ is zero for $\chi_n = \chi_\ell$, $\ell = 2, \ldots, n-1$ and for $\chi_n = \psi_\ell$, $\ell = 1, \ldots, n$. Hence, the function $S^r_{n, 1}(\{\chi\}; \{\psi\})$ has at least $4 n - 2$ zeros, and since $4 n - 2 > 3 n - 1$, we have $S^r_{n, 1}(\{\chi\}; \{\psi\}) = 0$.

Thus, we proved that
\[
\sum_{s=0}^2 F_n^{r + s}(\chi_1, \ldots, \chi_k + 2 \pi s / 3, \ldots, \chi_n; \{\psi\}) = 0,
\]
where the function $F_n^r(\{\chi\}; \{\psi\})$ is related to the partition function of the model via the equality (\ref{e:24}). In the same way we can show that
\[
\sum_{s=0}^2 F_n^{r + s}(\{\chi\}; \psi_1, \ldots, \psi_k - 2 \pi s / 3,  \ldots, \psi_n) = 0.
\]

\subsection*{Acknowledgment}
The work was partially supported by the Russian Foundation for Basic Research under grant \# 07-01-00234. We are grateful to H.~Rosengren for taking our attention to the paper \cite{Ros08} and for fruitful discussions.

\section*{Appendix}

After the substitutions $\lambda \to \lambda + \pi \tau / 2$ and $\varphi \to - \varphi - \pi / 3$, using the relation
\[
\theta_4(\varphi | p) = \rmi p^{1/4} \rme^{- \rmi \varphi} \theta_1(\varphi - \pi \tau / 2 | p),
\]
we transform the weights (\ref{e:1})--(\ref{e:4}) to the weights
\begin{align}
&\alpha_r(\varphi) = \alpha'_r(\varphi) = \zeta_r^{- 3 \varphi / 4 \pi} \, \frac{\theta_1(2 \pi / 3 + \varphi)}{\theta_1(2 \pi / 3)}, \label{e:31} \\
&\beta_r(\varphi) = \beta'_r(\varphi) = - \zeta_r^{1 / 2 +  3 \varphi / 4 \pi} \, \frac{\theta_1(\varphi)}{\theta_1(2 \pi / 3)}, \\
&\gamma_r(\varphi) = \rme^{\rmi \varphi} \, \frac{\zeta^{\varphi / 2 \pi}_r}{\zeta^{\varphi / 2 \pi}_{r + \overline 1}} \, \frac{\theta_1(\lambda + 2 \pi r / 3 - \varphi)}{\theta_1(\lambda + 2 \pi r / 3)}, \\
&\gamma'_r(\varphi) = \rme^{-\rmi \varphi} \, \frac{\zeta^{\varphi / 2 \pi}_r}{\zeta^{\varphi / 2 \pi}_{r - \overline 1}} \, \frac{\theta_1(\lambda + 2 \pi r / 3 + \varphi)}{\theta_1(\lambda + 2 \pi r / 3)}, \label{e:34}
\end{align}
where
\[
\zeta_r = \frac{\theta_1(\lambda + 2 \pi (r - 1) / 3) \theta_1(\lambda +  2 \pi (r + 1) / 3)}{\theta_1^2(\lambda + 2 \pi r / 3)}.
\]
These weights satisfy the Yang--Baxter equations of the form
\[
\sum_{t \in \bbZ_3} W_{r' t}^{r'' s''}(\varphi) W_{r \, s}^{r' t}(\varphi') W_{t \, s}^{s'' s'}(\varphi - \varphi') = \sum_{t \in \bbZ_3} W_{r' r}^{r'' t}(\varphi - \varphi') W_{t \, s'}^{r'' s''}(\varphi') W_{r\,  s}^{t \, s'}(\varphi)
\]
and the gauge transformations are again given by the formula (\ref{e:17}), where the functions $\Phi_r(\varphi)$ satisfy now the relation
\[
\Phi_r(\varphi - \varphi') = \frac{\Phi_r(\varphi)}{\Phi_r(\varphi')}.
\]
More explicitly the gauge transformations (\ref{e:17}) are described by the formulas
\begin{align*}
& \tilde \alpha_r(\varphi) = \frac{\Phi_{r - \overline 1}^{}(\varphi)  \Phi_{r + \overline 1}^{}(\varphi)}{\Phi_r^2 (\varphi)}  \alpha_r(\varphi),
&& \tilde \alpha'_r(\varphi) = \frac{\Phi_{r - \overline 1}^{}(\varphi)  \Phi_{r + \overline 1}^{}(\varphi)}{\Phi_r^2 (\varphi)} \alpha'_r(\varphi), \\[.5em]
& \tilde \beta_r(\varphi) = \frac{C_{r + \overline 1}}{C_{r - \overline 1}} \frac{\Phi_r^2(\varphi)}{\Phi_{r - \overline 1}^{-1}(\varphi) \Phi_{r + \overline 1}(\varphi)}   \beta_r(\varphi),
&& \tilde \beta'_r(\varphi) = \frac{C_{r - \overline 1}}{C_{r + \overline 1}} \frac{\Phi_r^2(\varphi)}{\Phi_{r - \overline 1}^{-1}(\varphi) \Phi_{r + \overline 1}(\varphi)}   \beta'_r(\varphi), \\[.5em]
& \tilde \gamma_r(\varphi) = \frac{\Phi_r^2(\varphi)}{\Phi_{r + \overline 1}^2(\varphi)}  \gamma_r(\varphi),
&& \tilde \gamma'_r(\varphi) = \frac{\Phi_r^2(\varphi)}{\Phi_{r - \overline 1}^2(\varphi)}  \gamma'_r(\varphi).
\end{align*}
Applying the gauge transformation (\ref{e:17}) with
\[
C_r = \rme^{\rmi \pi /2} \theta_1^{-1 / 2}(\lambda + 2 \pi r / 3), \qquad \Phi_r(\varphi) = \rme^{\rmi r \varphi / 2} \zeta_r^{-\varphi / 4 \pi}
\]
to the weights (\ref{e:31})--(\ref{e:34}) we obtain
\begin{align*}
&\tilde \alpha_r(\varphi) = \frac{\theta_1(2 \pi / 3 + \varphi)}{\theta_1(2 \pi / 3)}, &&\tilde \alpha'_r(\varphi) = \frac{\theta_1(2 \pi / 3 + \varphi)}{\theta_1(2 \pi / 3)}, \\
&\tilde \beta_r(\varphi) = \frac{\theta_1(\lambda + 2 \pi (r - 1) / 3)}{\theta_1(\lambda + 2 \pi r / 3)} \, \frac{\theta_1(\varphi)}{\theta_1(2 \pi / 3)},  &&\tilde \beta'_r(\varphi) =  \frac{\theta_1(\lambda + 2 \pi (r + 1) / 3)}{\theta_1(\lambda + 2 \pi r / 3)} \, \frac{\theta_1(\varphi)}{\theta_1(2 \pi / 3)}, \\
&\tilde \gamma_r(\varphi) = \frac{\theta_1(\lambda + 2 \pi r / 3 - \varphi)}{\theta_1(\lambda + 2 \pi r / 3)}, &&\tilde \gamma'_r(\varphi) = \frac{\theta_1(\lambda + 2 \pi r / 3 + \varphi)}{\theta_1(\lambda + 2 \pi r / 3)}.
\end{align*}
These weights coincide with the weights used in the paper \cite{Ros08}.

\end{document}